\documentclass[aps, twocolumn,amsmath, ammsymb]{revtex4-1}
\usepackage{epsfig}
\usepackage[usenames]{color}
\usepackage{bm}

\newcommand{\nn}{\nonumber}

\newcommand{\bs}{\mathbf{s}}

\newcommand{\bB}{\mathbf{B}}

\newcommand{\cH}{{\cal H}}

\newcommand{\dg}{\dagger}
\newcommand{\bS}{\mathbf{S}}
\newcommand{\bmu} {\bm \mu}
\newcommand{\bchi} {\bm \chi}

\newcommand{\be}{\begin{eqnarray}}
\newcommand{\ee}{\end{eqnarray}}
\newcommand{\la}{\langle}
\newcommand{\ra}{\rangle}

\newcommand{\da}{\downarrow}
\newcommand{\ua}{\uparrow}

\begin{document}

\title{Universality in higher order spin noise spectroscopy}

\author{ Fuxiang Li$^{1,2}$, and N. A. Sinitsyn$^{2,*}$}
\address{$^1$ Center for Nonlinear Studies, Los Alamos National Laboratory,  Los Alamos, NM 87545 USA}
\address{$^2$ Theoretical Division, Los Alamos National Laboratory,   Los Alamos, NM 87545 USA}
%\email{nsinitsyn@lanl.gov}

\date{\today,\now}

\begin{abstract}
Higher order time-correlators of spin fluctuations reveal  considerable information about spin interactions. 
We argue that in a broad class of spin systems one can justify a phenomenological approach to explore such correlators. We predict that 
the 3rd and 4th order spin cumulants are described by a universal function that can be parametrized by a small set of parameters. We show that the fluctuation theorem constrains this function so that such correlators are fully determined by lowest nonlinear corrections to the free energy and the mean and variance of microscopic spin currents. We also provide an example of microscopic calculations for conduction electrons.
 
%The theory of higher than 2nd order spin correlations is in early stage of development.
 
%The 
%fluctuation theorem is applied to derive a symmetry in the Hamiltonian, and this symmetry relates the higher order interacting terms in Hamiltonian to the non-equilibrium, lower order terms.  Through this symmetry, to obtain the fourth order cumulant of spin fluctuations, one only needs to calculate the second order non-quilibrium  cumulants.  A microscopic model of electron system scattering with random magnetic impurities is provided. 
\end{abstract}

\date{\today}
\maketitle

%{\it Introduction.} 
Spin noise spectroscopy (SNS) \cite{aleksandrov1981magnetic,crooker2004spectroscopy} is a  quickly 
evolving interdisciplinary field of research. It explores spin  interactions by tracing dynamics of spontaneous spin fluctuations at or near the thermodynamic equilibrium without the need to intentionally polarize  spins.   The SNS  has been successfully applied to  semiconductors \cite{oestreich2005spin, muller2008spin, crooker2009spin}, quantum dots \cite{crooker2010spin, li2012intrinsic, yang2014two}, hot and ultra-cold  atomic gases \cite{crooker2004spectroscopy,katsoprinakis2007measurement, zapasskii2013optical}. 

So far the SNS has been focused on studies of the 2nd order spin correlator, $\la S_z(t) S_z(0)\ra$, of a local time-dependent spin polarization $S_z(t)$,  or rather on its Fourier transform called the spin noise power spectrum:
\begin{equation} 
C_2(\omega) \equiv \langle |S_z(\omega)|^2 \rangle, 
\label{noiseP1}
\end{equation}
where $S_z(\omega) = \frac{1}{\sqrt{T_m}} \int_0^{T_m} dt \, S_z(t) e^{i\omega t}$,  and averaging is over many repeated time intervals of duration $T_m$.
The information content of the correlator (\ref{noiseP1}) is strongly restricted. Hence, one of the promising future directions to extend the SNS is to measure higher order spin correlators \cite{mukamel-2D, Bechtold2015, ubbelohde2012measurement, branczyk2014crossing}, the most accessible of which are the 3rd and 4th order ones:
\be
C_3 \equiv \la S_z(\omega_1) S_z(\omega_2) S_z(-\omega_1-\omega_2) \ra,
\label{c3-1}
\ee
\be
C_4 \equiv \la |S_z(\omega_1)|^2 |S_z(\omega_2)|^2 \ra - \la |S_z(\omega_1)|^2 \ra \la |S_z(\omega_2)|^2 \ra, 
\label{c4-1}
\ee
which depend on two frequencies, $\omega_1$ and $\omega_2$. 

Unlike the noise power (\ref{noiseP1}) that describes the spectral frequency weights, the bi-spectra (\ref{c3-1})-(\ref{c4-1}) tell how different frequencies `talk' to each other. These correlators are sensitive to many-body interactions  \cite{li2013higher} and quantum effects \cite{alex-2}, suggesting that their studies by the SNS can reveal essentially new information about correlated spin systems.
As  the noise of a single spin in a quantum dot \cite{dahbashi2014optical} and the noise of only a few hundreds of spins of conduction electrons in a 2D electron gas \cite{2D-noise} have already been studied experimentally, the goal to obtain the 3rd and 4th order spin correlators experimentally becomes achievable \cite{alex-2}. However, very little is known about properties of $C_3$ and $C_4$ in basic systems studied by the SNS, such as conduction electrons and atomic gases, e.g., about how  (\ref{c3-1})-(\ref{c4-1}) are influenced by the Pauli principle, scatterings, spin orbit coupling, and external magnetic field. There have been no quantum mechanically justified studies of such correlators in interacting electronic systems. %, apart from complex diagrammactic techniques that have not been attempted yet.

An  important observation made throughout all known SNS applications is that the spin noise often shows well recognized patterns. For example, the noise power spectrum often consists of one or several peaks having the Lorentzian shape. The position and the width of such a peak determine useful parameters:  the g-factor of the spin resonance and its life time \cite{crooker2004spectroscopy}. This universality is not a coincidence and it is well understood: the Lorentzian shape of a peak indicates exponential relaxation in time that happens due to  fast uncorrelated interactions. For example, conduction electrons that experience fast fluctuations of the spin-orbit field usually demonstrate a Lorentzian shape of the spin noise power spectrum \cite{li2013nonequilibrium}. In fact, the commonly used Bloch equation and various relaxation time approximations are justified by exactly this type of universality. 

In this letter, we argue that a similar universality exists on the level of higher order spin correlators, namely,  under the conditions that the noise power spectrum is Lorentzian, the 3rd and 4th spin correlators  can also be parametrized by a small set of parameters with a clear physical meaning. In addition, we show that, as a consequence of the fluctuation theorem,  such parameters are not independent when the system is probed at or near the thermodynamic equilibrium.  

%As an example, we will derive these correlators for a specific model of conduction electrons interacting with random static magnetic impurities. 

%{\it Higher order terms in Hamiltonian.} 
A Lorentzian peak in the noise power spectrum indicates that the dynamics of the spin fluctuation follows the  Bloch-Langevin equation: % that describes dynamics of a spin fluctuation in  a mesoscopic region of a paramagnetic material, e.g, in conduction electrons of a semiconductor:
\be
\dot{\bf S} = {\bf B} \times {\bf S} - \gamma {\bf S} +{\bm \eta}(t), 
\label{bloch1}
\ee
where $\gamma$ is the relaxation rate, and ${\bm \eta}$ is the noise term:
\be
\la {\bm \eta} \ra =0, \,\,\, \la { \eta_i}(t) {\eta_j}(t') \ra = 2D_{ij} \delta(t-t'); \,\,  i,j=x,y,z. 
\ee
 For simplicity, we assumed an isotropic relaxation rate, and we absorbed  the g-factor in the definition of the magnetic field ${\bf B}$. Although phenomenological, Eq.~(\ref{bloch1}) has been highly successful to describe diverse spin systems, including conduction electrons, nuclear spins, warm atomic gases, and hole spins of quantum dots \cite{crooker2004spectroscopy,oestreich2005spin,li2012intrinsic,katsoprinakis2007measurement}. 
For small spontaneous fluctuations in a region with $N \gg 1$  spins, we can disregard the dependence of the relaxation rate $\gamma$ on ${\bf S}$. The probability of a fluctuation  ${\bf S} \ne 0$ at the thermodynamic equilibrium is determined by the free energy $F({\bf S})$ of spins in the region at known polarization ${\bS}$:
\be
P({\bf S}) \sim e^{-\beta F({\bf S})}, \quad  F= \frac{a_0}{2} \bS^2 + \frac{b_0}{4} \bS^4 + ...,
\label{prob1}
\ee
where $\beta =1/k_BT$. Hence,  $D_{ij}$ are not always independent parameters.  In order to reproduce the Gaussian part in (\ref{prob1}), one should set $D_{ij}=D_0\delta_{ij}$, where $D_0=\gamma/(\beta a_0)$.

We will assume that the magnetic field is applied along the $y$-axis, which is transverse to the measurement $z$-axis. 
The noise power spectrum produced by  Eq.~(\ref{bloch1}) is then well known. It consists of two Lorentzian peaks \cite{li2013nonequilibrium}:
\be
C_2(\omega)= \sum _{s=\pm} \frac{D_0}{(\omega-s\omega_L)^2+\gamma^2}, \quad \omega_L\equiv  |{\bf B}|.
\label{noiseP2}
\ee 
Equation (\ref{bloch1}) with constant parameters $\gamma$ and $D_0$ predicts zero values for all higher than 2nd order spin cumulants. 

Our first observation is that it is straightforward to generalize Eq.~(\ref{bloch1}) to include nonlinear effects. Correlators (\ref{c3-1})-(\ref{c4-1}) must follow then from higher order corrections to parameters that, for an isotropic system, read
\begin{eqnarray}
\label{gamma2}
\gamma&=&\gamma (\bS) = \gamma_0+\gamma_2 \bS^2+ O(\bS^4), \\
\label{d2}
D_{ij}&=&(D_0+D_1 \bS^2)\delta_{ij} + D_{2}(S_i S_j - \bS^2 \delta_{ij}) + O(\bS^4).
\end{eqnarray}
Here $D_1$ corresponds to renormalization of the noise part related to dissipative spin relaxation, and the term with $D_2$ describes angular diffusion of a spin fluctuation without relaxation of its absolute value \cite{supp}. 
%NOTE: CAN WE CONNECT $D_{ij}$ to $D_1$ and $D_2$. WHY $D_{ij}$ IS NOT A SINGLE CONSTANT? E.G. CAN WE INTRODUCE PARAMETERS $D_1$ and $D_2$ HERE AND SAY THAT ONE DESCRIBES FLUCTUATIONS OF THE ABSOLUTE VALUE AND ANOTHER IS RELATED TO THE DIFFUSION OF THE SPIN ANGLE, OR SOMETHING LIKE THIS - IT WOULD BE VERY USEFUL.

%Similarly, the 2nd order noise correlator can be the function of the fluctuation size, i.e., $D=D(\bS)$. 

Microscopic calculations of the leading corrections  to the nonlinear relaxation rate  $\gamma$ and the 2nd order noise correlator $D_{ij}$ can be done within the approach developed in \cite{li2013higher}. The biggest complication, however, is that the knowledge of $\gamma (\bS) $ and $D_{ij}(\bS)$ is insufficient to determine $C_3$ and $C_4$ because the noise term ${\bm \eta}$ in (\ref{bloch1}) can no longer  be considered Gaussian, while a microscopic quantum theory to obtain non-Gaussian statistics of ${\bm \eta}$  is generally missing. Below we show, and this will be one of our key results, that this problem can be avoided for fluctuations at the thermodynamic equilibrium because non-Gaussian correlations of ${\bm \eta}$ are then uniquely constrained, i.e. they can be derived without additional microscopic calculations.

To address this problem, we note that, for a mesoscopic system, one can choose the time step $\delta t$ such that the number of spin flips in the system is large but still much smaller than the typical size of the spin fluctuation ${\bf S}$. Let $P[\delta\bS|\bS]$
be the probability of observing the change of the spin polarization by the amount $\delta \bS$ during $\delta t$ given that initially the spin fluctuation size is $\bS$.
The law of large numbers guarantees that cumulants of $\delta \bS$ grow  linearly with $\delta t$ as far as $\la \delta \bS \ra \ll \bS$ \cite{pilgram2003stochastic}. This fact can be expressed by introducing the cumulant generating function $\cH(\bchi, \bS)$:
\be
P[\delta \bS|\bS]  = \int d \bchi e^{ i  \bchi \cdot \delta \bS}  e^{\delta t \cH(\bchi, \bS)} .
\label{prob-lin}
\ee
\begin{figure}[!htb]
\scalebox{0.36}[0.36]{\includegraphics{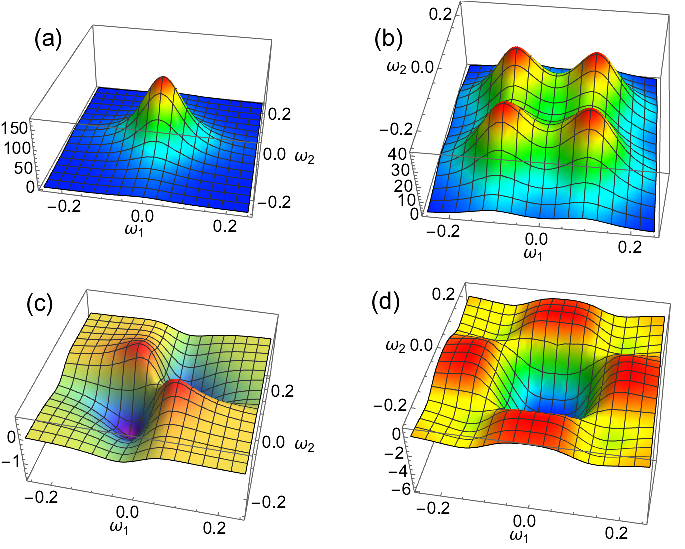}}
%\scalebox{0.17}[0.18]{\includegraphics{contour2.pdf}}
%\hspace{-2mm}\vspace{-4mm}
\caption {Fourth order cumulants in the presence of a magnetic field for different values of parameters: (a) $D_1= 1$, $D_2 = 0$, $\omega_L=0.0$;  (b) $D_1= 1$, $D_2 = 0$, $\omega_L=0.1$; (c) $D_1= 0$, $D_2 = 1$, $\omega_L=0$; (d) $D_1= 0$, $D_2 = 1$, $\omega_L=0.1$.    Other parameters are: $\gamma_0=0.1$,  $a_0=1$, $b_0=0$, $\beta=1$ and $\gamma_2=0$.}
\label{fig:c4}
\end{figure}

Next, we note that fluctuations near the thermodynamic equilibrium should satisfy the detailed balance constraints, so that probabilities of spin polarization changes by $\delta \bS$ and $-\delta \bS$ are related by \cite{Bertini2015Macro, crooks1999entropy} 
\begin{equation}
\frac{P[\delta {\bf S}|\bS ]}{\tilde{P}[-\delta {\bf S}|\bS +\delta \bS]} =e^{\beta (F(\bS)-F(\bS+\delta \bS))} \approx e^{- \beta \delta {\bf S \cdot}{\bm \mu}}, \,\,\,  {\bm \mu}=\frac{\partial F}{\partial \bS},
\label{work2}
\end{equation}
where tilde sign in ${\tilde{P}[-\delta {\bf S}|\bS +\delta \bS]}$ means the probability at time-reversed values of model parameters.  In combination with (\ref{prob-lin}), Eq.~(\ref{work2}) leads to the fluctuation relation (which is a special case of Eq.~(3.2.51) in \cite{stratonovich1992}):
\be
\cH(\bchi, -\bS, -\bB)=\cH(\bchi + i\beta \bmu, \bS, \bB),
\label{eq:Hsym2}
\ee
where $\cH$ is defined in (\ref{prob-lin}).
%\be
%{\cal Z} (\bchi) = e^{\delta t \cH(\bchi, \bS)} \label{eq:cH}
%\ee
%in which, $\cH$ is the term in the effective Lagrangian that describes fluctuations of the variable $\bS$, and is, in general, a function of variable $\bS$ and external magnetic field $\bB_e$. 
Next, we recall that if $\cH(\bchi, \bS)$ is known, arbitrary spin correlator can be derived by the method of the stochastic path integral \cite{li2013higher, pilgram2003stochastic}. The latter is  the sum, $Z$, over all possible stochastic trajectories, discretized in time-steps $\delta t$, of random variables ${\delta \bS}(t)$ and $\bS(t)$ weighted by probabilities (\ref{prob-lin}) and delta-functions $\delta \left(\dot{\bS}-\delta \bS/\delta t \right)$ at each elementary time interval. Integrals over ${\delta \bS}(t)$  can be performed explicitly with an expense of introducing a conjugated to  $ \bS(t)$ variable $\bchi (t)$.  
%Specifically, the partition function $Z= \int D \bS(t)  \int D {\delta \bS}  P[\delta\bS | \bS] {\bm \delta} (\delta t \dot{\bS}(t)  - \delta \bS )$. Substituting Eq.~(\ref{prob-lin}) into it and performing the integral over $\delta \bS$, we arrive at 
%Derivation of the path integral is straightforward  H
Following  \cite{li2013higher, pilgram2003stochastic}, we find:
\be
Z=\int  D \bS(t) \int D \bchi(t) \ e^{\int dt \left( i \bchi \dot{\bS} +\cH(\bchi, \bS) \right)}.
\label{path}
\ee
 In order  to derive an  n-th order spin correlator in the $N\gg 1$ limit, it is enough to keep only terms up to the n-th power of variables $\bchi$, $\bS$  in $\cH(\bchi, \bS)$. Assuming an isotropic paramagnetic system, the most general form of $\cH(\bchi, \bS)$ up to the 4th power of variables is
\begin{equation}
\cH =  i\gamma (\bS) \bchi \cdot \bS + i \bchi \cdot (\bS\times \bB)-\hat{D}({\bS},\bchi) + i D_3  {\bm \chi}^2 (\bS\cdot \bchi)+ D_4 {\bm \chi}^4, % \hat{D}_3 ( {\bm \chi} )
\label{hh4}
\end{equation}
where 
\begin{equation}
\hat{D}(\bS,\bchi) = D_0  {\bm \chi}^2 + (D_1-D_2) \bS^2  {\bm \chi}^2 +D_2  (\bS\cdot   {\bm \chi})^2.
\end{equation}
Here constants $D_0,D_1,D_2$ have the same physical meaning as in (\ref{d2}), and $\gamma(\bS)$ is given by the two first terms in (\ref{gamma2}).
Importantly,  applying the symmetry (\ref{eq:Hsym2}) to (\ref{hh4}), we find the standard fluctuation-dissipation theorem prediction $D_0=\gamma_0/(\beta a_0)$, as well as the additional constraints relating higher order coefficients:
\be
D_4 = \frac{ \beta b_0   D_0 + \beta a_0   D_1 -\gamma_2 }{\beta^3 a_0^3},  \, \, D_3 = -2 \beta a_0 D_4.
\label{eq:d3}
\ee
%where $a=\beta a_0, ~~ b=\beta b_0 $. 
%The terms of the 2nd order in ${\bchi}$  in (\ref{hh4})  can be identified with  parameters in (\ref{d2}) while
Equation~(\ref{hh4}) with constraints (\ref{eq:d3}) are the central results of this article. They show that the information about up to the 4th order spin correlators is contained in a simple function $\cH$ that depends on a small set of parameters. Moreover, note that terms of higher than 2nd power of $\bchi$ in (\ref{hh4}) characterize higher order correlators of the noise term 
${\bm \eta}$ in (\ref{bloch1}). The fact that corresponding coefficients $D_3$ and $D_4$ can be written in terms of coefficients at nonlinear corrections to the 2nd order correlators of ${\bm \eta}$ and the quartic correction to the free energy in (\ref{prob1}) considerably simplifies the goal of their microscopic calculation. % of $C_3$ and $C_4$ because the theory 

%, while $D_0$ is constrained by the fluctuation-dissipation theorem

%NOTE: WE SHOULD ALSO MENTION THAT $D_0$ IS RELATED TO $\lambda$. IDEALLY, WE SHOULD EXPLAIN THIS EARLY IN RELATION TO EQ. (4) AND HERE JUST MENTION THIS AGAIN. PLEASE THINK HOW TO REWRITE EQ. 4 TO INTRODUCE PARAMETER $D_0$ INSTEAD OF $D$.

% Moreover, parameters $D_0$, $D_3$ and $D_4$ are not independent, namely, . 

Remaining  parameters depend on microscopic spin dynamics. Nevertheless, the  universality that we found allows us to look at  {\it possible patterns} of 3rd and 4th correlators.
They can  be calculated by switching to the frequency domain in the action  of the path integral and  treating 4th order terms in (\ref{hh4}) as a small perturbation:
%At the thermodynamic equilibrium, the 3rd order correlator is zero due to the time reversal symmetry, while the 4 order correlator is given by expression:
\be
C_4(\omega_1, \omega_2) = \la |S_z(\omega_1)|^2 |S_z(\omega_2)|^2 {\cal{R}}_{4}\ra_0
\ee
where ${\cal{R}}_4 = \int dt {\cal H}_4$ and ${\cal H}_4$ is the quartic part of the Hamiltonian  in Eq.~(\ref{hh4}), and  $\la...\ra_0$ means that the average is taken over the quadratic action in the path integral. 
After applying the Wick's rule, we find a relatively complex  expression that we provide in supplementary material \cite{supp}.
In Fig.~\ref{fig:c4}, we show examples of the obtained 4th correlator shapes at different values of independent parameters, including the magnetic field.

%{ \it Example of electron system scattering with magnetic impurities.}

Analogously, we can explore the form of the 3rd order correlator. It becomes nonzero at nonequilibrium conditions, for example, when a finite spin density is induced in conduction electors by a resonant optical pumping. We will consider the limit of a weak 
intensity of the pumping beam so that spin generation happens in uncorrelated events that have a Poisson distribution, which contributes to the Hamiltonian in the action of Eq.~(\ref{path}) with a term \cite{pilgram2003stochastic}:
\be
\cH_p = k_p \Big( e^{-i\chi_z } - 1\Big), 
\ee
where $k_{p}$ is the generation rate of spin polarization by an optical beam. Since the 3rd order correlator is zero at $k_p=0$, and since we are interested in the linear,  in $k_p$, response contribution, we can disregard the effect of small renormalization of all other terms in the action of the path integral (\ref{path}) on $C_3$. 
The saddle point equations with the total Hamiltonian $\cH_T=\cH+\cH_p$ read:
\be
\delta \cH_T/\delta \chi_i= 0,  ~{\rm and }~~ \delta \cH_T/\delta S_i =0, ~~ {\rm with }~ i = x, z. \nn
\ee
They have a solution $\bar{\chi}_i=0$ and
\be
\bar{S}_z  = \frac{k_p \gamma_0 }{\gamma_0^2 + \omega_L^2}, ~~
\bar{S}_x  = \frac{ k_p \omega_L}{\gamma_0^2 + \omega_L^2} .
\label{avg1}
\ee
%where $\omega_L$ is the magnitude external magnetic field. 

%This term will produce a nonzero value of $S_z$ and $S_x$ (if we assume that the external magnetic field directs along $\hat{y}$.), 
%which correspond to the  saddle point solution of the following equation of motion:

%The higher order spin correlators can be obtained by considering the action near the saddle point beyongd the classical limit and introduce finite values $\bchi$ and $\bS$ to describe the deviation.  Since we are interested only in the third order cumulant, it is sufficient to present the Hamiltonian in the action up to third order in total powers of $\bchi$ and $\bS$:

 \begin{figure}[!htb]
\scalebox{0.23}[0.23]{\includegraphics{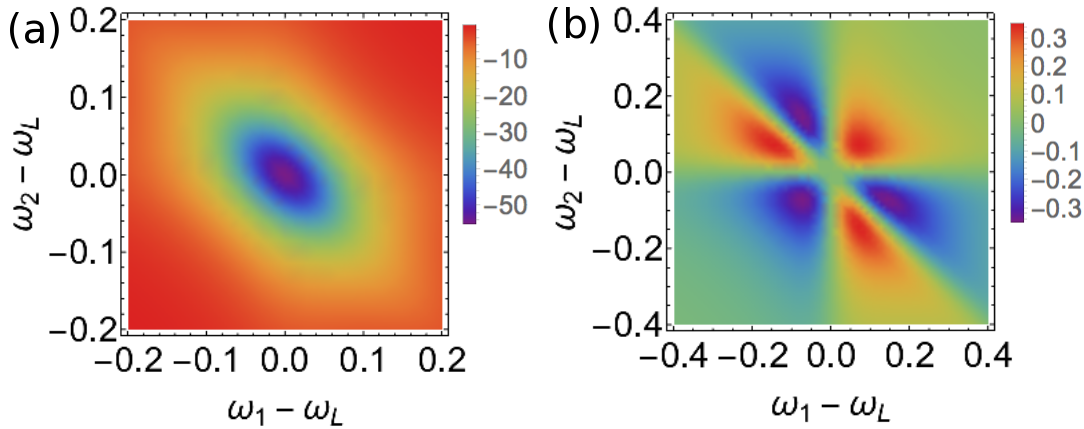}}
%\scalebox{0.17}[0.18]{\includegraphics{contour2.pdf}}
%\hspace{-2mm}\vspace{-4mm}
\caption{The real (a) and imaginary (b) parts of third order cumulant in the regime of a continuous spin pumping along $z$ axis with $k_p=0.3$. Parameters are: $D_1=D_2=1$, $\gamma_0=0.1$, $\omega_L=1$, $a_0=1$, $b_0=0$, $\beta=1$ and $\gamma_2=0$.}
\label{fig:c3}
\end{figure} 
By expanding the action in powers of small fluctuations $\delta \bS$ and $\bchi$ from the steady state, the 3rd order  in $\delta \bS$ and $\bchi$ part of the Hamiltonian in the path integral reads:
%\begin{widetext}
\be
\cH_{T,3} &=& -\frac{1}{2} k_p \chi_z^2 \delta S_z - 2 (D_1 - D_2) (  \bar{\bS}  \delta  \bS) \bchi^2    \nn \\
&&  -2 D_2 (\bar{\bS} \cdot \bchi) (\bchi \cdot  \delta \bS)  +  iD_3  (\bar{\bS} \cdot \bchi) \bchi^2
\nn\\
 && +i\gamma_2 ((\bar{\bS} \cdot \bchi) (\delta \bS)^2  + 2 (\delta  \bS \cdot \bar{\bS}) (\bchi \cdot \bS) ).   
\ee
%\end{widetext}
We find that corresponding correlator $C_3$ is generally complex valued. In Fig.~\ref{fig:c3}(a) and (b) we plot, respectively,  the real and imaginary parts of a typical pattern of $C_3(\omega_1, \omega_2)$.

%{\it Example.} 
As a demonstration of a microscopic estimate of independent parameters, we consider a  model of conduction electrons, for which spin relaxation and fluctuations are caused by scatterings on weak randomly distributed static magnetic impurities.  
We assume that  energy-momentum degrees of freedom of electrons equilibrate to the Fermi-Dirac distribution at the ambient temperature  very quickly, while the spin degrees of freedom equilibrate at a much slower rate.  
%Therefore, we can describe the system by a density matrix that is diagonal in the momentum $\bk$ space but is not diagonal in the spin space for each $\bk$ state.
 For an open isotropic  region around the studied spot, we can introduce a local instantaneous vector chemical potential $\bmu(t)$ to describe the single particle density matrix of electrons:
\be
\hat{\rho}(\epsilon)&=&\frac{1}{1+e^{\beta (\epsilon-\bm{\mu}\cdot \hat{\bm \sigma}/2)}}  \label{eq:densitymatrix},
\ee
% & \equiv &  \rho_{0,\bk} + {\bm \rho}_{\bk} \cdot  {\bm \sigma} 
where $\bm{\hat\sigma}$ is the vector made of Pauli matrices acting in the spin space. The  potential ${\bm \mu}$ is a slow variable that changes at the spin relaxation time scale. We assume that the observation region is much smaller than the spin diffusion length. The average spin polarization density is then given by $\bS(t) = d_s \int d\epsilon {\rm Tr} [{\bm{\hat \sigma}} \hat{\rho}]/2 = d_s {\bm \mu}(t)/2$, where $d_s$ is the density of states per unit energy.
Fast transitions of electronic spins through the observation region and quantum measurement effects will induce fast noise $\xi(t)$ in the measured Faraday rotation angle $\theta_F$ of the probe beam.  However, averaging the signal over an interval $\delta t$ that is smaller but comparable to spin relaxation time, one would find that $\la \theta_F \ra \sim S_z \delta t $, i.e. the information about $S_z$ accumulates with time and dominates 
the measured correlations at spin relaxation time scale, while $\xi(t)$ contributes to the background noise and the high frequency tail of the spectrum, which we will not study here.

For  a fermionic system with the density matrix (\ref{eq:densitymatrix}), the free energy is a quadratic function of a parameter ${\bS}$: $F=\int \bmu d \bS = {\bS}^2/(d_s)$. 
Let $\hat{\Psi } \equiv (\hat{a}_{\ua}, \hat{a}_{\da})$,  where $\hat{a}_{\ua, \da}$  are the annihilation operators of electrons in two degenerate eigenstates of the Hamiltonian that includes non-magnetic disorder. Weak interaction with magnetic impurities couples these states so that, in the Dirac picture, annihilation operators evolve with time: 
\be
\hat{\Psi}(\delta t) =( \cos\theta + i {\bf m}\cdot {\bm{\hat\sigma}}  \sin\theta ) \hat{\Psi}(0), \label{scatt}
\ee
where  $\theta$ is a random parameter of the evolution matrix, such that averaging over all scattering channels $\la \theta^2 \ra \sim \delta t$ and, for each scattering channel, ${\bf m}$ is a randomly directed unit vector. 
Let  $\hat{\bf s}=\frac{1}{2} \hat{\Psi}^{\dagger} {\bm \hat{\bm \sigma}} \hat{\Psi}$ be the spin operator in the subspace of the two states. 
As we showed before, to determine parameters of the path integral action we need to know only up to 2nd order correlators of the spin change during a small time interval $\delta t$, which can be obtained by a simplified procedure, known to work only for such lowest order correlators \cite{stratonovich1992}.  Namely, we introduce the operator of the change of the spin: $ \delta\hat{\bs} \equiv  \hat{\bf s}(\delta t)-\hat{\bf s}(0)$.  Explicitly:
 \be
\delta \hat{\bs} &=& \frac{1}{2} \sin^2\theta \Big[ \Psi^{\dg} ({\bf m}\cdot {\hat{\bm \sigma}})\hat{\bm \sigma} ({\bf m}\cdot {\hat{\bm \sigma}})\Psi  - \Psi^{\dg} \hat{\bm \sigma} \Psi \Big]  \nn\\
&&+\frac{i}{2} \sin\theta \cos\theta \Big[ \Psi^{\dg}  \hat{\bm \sigma}  ({\bf m}\cdot {\hat{\bm \sigma}}) \Psi - \Psi^{\dg}  ({\bf m}\cdot {\hat{\bm \sigma}}) \hat{\bm \sigma} \Psi \Big].  \nn
 \label{eq:deltas}
\ee
Then the average change of the spin density and its variance are given by
$\la \delta \bS \ra = d_s \int d\epsilon  {\rm Tr}\Big[ \hat{\rho} \delta\hat{\bs} \Big]$ and $\la \delta S_{\alpha} \delta S_{\beta} \ra = d_s \int d \epsilon  {\rm Tr}\Big[ \hat{\rho} \{ \delta\hat{s}_{\alpha}, \delta\hat{s}_{\beta} \}/2 \Big],$ where curly brackets denote the anticommutator.
 % If we assume that scattering is weak, then $|t_{\bf k k'}|^2= |t_{\bf k' k}|^2\ll 1$, $|r_{\bf k k'}|^2=1-|t_{\bf k k'}|^2$. The scattering probability $|r_{\bk, \bk'}|^2$ scales linearly with time for two stakes $\bk$ and $\bk'$ with the same energy. Therefore it is convenient to introduce the scattering rate $\omega_{\bk, \bk'} = |r_{\bk, \bk'}|^2 /t$, which will be determined by the Fermi's golden rule.  The spin current operator, $\hat{\bJ}_{\bk \rar \bk'}$, due to such scattering is defined by:$\int_0^t dt' \hat{\bJ}_{\bk \rar \bk'}(t') =\delta\hat{\bs}_{\bk\rar \bk'}\equiv  \hat{\bf s}_{\bf k}(t)-\hat{\bf s}_{\bf k}(0) $.  Explicitly, the spin flux operator $ \delta\hat{\bs}_{\bk\rar \bk'} $:
% Let us define the change of . 
 %Using (\ref{scatt}):
%\be
%\delta \hat{\bs}_{\bk\rar\bk'} &=& \frac{|r_{\bk, \bk'}|^2}{2}  \Big[ \Psi^{\dg}_{\bk'} ({\bf m}\cdot {\hat{\bm \sigma}})\hat{\bm \sigma} ({\bf m}\cdot {\hat{\bm \sigma}})\Psi_{\bk'}  - \Psi_{\bk}^{\dg} \hat{\bm \sigma} \Psi_{\bk} \Big]  \nn\\
%&&+\frac{i}{2} t_{\bk, \bk} r_{\bk, \bk'} \Big[ \Psi_{\bk}^{\dg}  \hat{\bm \sigma}  ({\bf m}\cdot {\hat{\bm \sigma}}) \Psi_{\bk'} - \Psi_{\bk'}^{\dg}  ({\bf m}\cdot {\hat{\bm \sigma}}) \hat{\bm \sigma} \Psi_{\bk} \Big] \nn 
%\\ \label{eq:deltas}
%\ee
Using the density matrix (\ref{eq:densitymatrix}) in the secondary quantized form, we obtain spin fluctuations due to scatterings between particular degenerate spin states. 
%To average over scattering matrix elements, we assume the weak scattering limit $\theta \ll 1$ with $\la \theta^2 \ra \sim \delta t$ and a uniform distribution of  the direction of unit vector ${\bf m}$ (\ref{scatt}). 
Finally, we assume that all scattering channels are independent during time $\delta t$,  average over parameters $\theta$ and ${\bf m}$, and integrate over  $\epsilon$. %Moreover, we identify $\la \delta {\bm \mu} \ra =\la \delta {\bs}\ra/d_s $ e.t.c. 
We find 
  $\la \delta \bS \ra/\delta t\equiv - \gamma_0 \bS=  - \frac{4}{3} \bS  \la \theta^2 \ra/\delta t$, and $$ \la \delta S_{\alpha} \delta S_{\beta} \ra  = \delta_{\alpha \beta} \frac{4}{3} \la \theta^2 \ra  \Big( \frac{d_s }{ \beta }+ \frac{ \beta }{3 d_s } \bS^2 \Big).$$  % $\la \delta { \mu}_{\alpha} \delta \mu_{\beta} \ra = XX$. [NOTE: PLEASE FILL IN THE MISSING COEFFICIENTS X and XX]. 
  Comparing  with Eq.~(\ref{bloch1}) and (\ref{d2}), we identify
\be
D_0  =\gamma_0 d_s/(2 \beta), ~~ D_1 = \gamma_0 \beta/(6 d_s), ~~ D_2=0.
\ee 
In this case, the coefficient  in the free energy $a_0= 2/d_s$. Therefore, we recover the fluctuation-dissipation relation $D_0=\frac{\gamma_0}{a_0 \beta}$.
Using Eq.~(\ref{eq:d3}), and considering the fact that  $b_0=0$ and $\gamma_2=0$ we find:
\be
D_4= d_s \gamma_0/(24 \beta ),   ~~~
D_3= -   \gamma_0/6 .
\ee
The resulting Hamiltonian in the path integral for this model  reads: 
\be
\cH &=& i\gamma_0 \bchi \cdot \bS  + i \bchi\cdot (\bS \times \bB)-  \frac{\gamma_0 d_s}{2 \beta} \bchi^2 - \frac{\gamma_0 \beta}{6 d_s } \bS^2 \bchi^2  \nn\\
&&~ -i \frac{\gamma_0}{6} (\bchi \cdot \bS)\bchi^2 + \frac{d_s \gamma_0 }{ 24 \beta } \bchi^4. 
\ee
Thus, this model corresponds to the case  with  $D_2=0$, which is illustrated in Fig.~\ref{fig:c4}(a) and (b).

One additional consequence of this result is the prediction  that the 4th correlator of a Fermi liquid scales linearly with temperature. For example, the $D_1$-term  contribution to the 4th cumulant, according to the Wick's theorem, is of the order of $C_4^{(1)} \sim D_1 \la \bS \bchi \ra_0^2  \times \la \bS \bS \ra_0^2 \sim  \beta \times 1 \times \frac{1}{\beta^2} = \frac{1}{\beta}$, where we used the fact that,  for  the second order correlators, the temperature dependence scales as $\la \bS \bchi \ra_0 \sim \beta^0$ and $\la \bS \bS \ra_0 \sim \beta^{-1}$.

In conclusion, we  developed a phenomenological approach that extends the Bloch-Langevin equation for spin dynamics to include the 3rd and 4th order spin correlations. 
This approach is justified by the law of large numbers and the higher order fluctuation relations. 
 Our theory should be applicable practically to all spin systems, near the thermodynamics equilibrium, that exhibit Lorentzian peaks in the spin noise power spectrum. Such cases are ubiquitous. Therefore, although the microscopic theory of higher order spin correlations is at the early stage of development, our results make a valuable insight into the possible forms of such correlators and  their dependence on temperature and optical spin pumping. 
 
 %We believe that our work will allow experimentalists to foresee the most likely forms of higher order spin correlators that can encounter in numerous systems: quantum dots, conduction electrons, atomic vapors. Moreover, constraints that we discovered strongly reduce the complexity of  microscopic calculations of such correlators, as we showed explicitly in a specific example of a Fermi liquid interacting with a weak magnetic disorder. 

% that the Fluctuation theorem leads to a symmetry of the Hamiltonian, which connect the higher order interacting coefficients with those determining the second order spin cumulants in non-equilibrium. This connection makes it possible to calculate the higher order spin cumulants from microscopic model, without resorting to other, if exiting, more sophisticated techniques.
{\it Acknowledgments}. Authors thank A. V. Andreev for useful discussions. 
The work
was carried out under the auspices of the National Nuclear
Security Administration of the U.S. Department of Energy at Los
Alamos National Laboratory under Contract No. DE-AC52-06NA25396. Authors also thank the support from  LDRD program at LANL.

%%%%%%%%%%%%%%%%%%%%%%%%%%%%%%%%%%%%%%%%%%%%%%%%%%%%%

\bibliographystyle{apsrev4-1}
%\bibliography{bib_sns}

\newpage 
\newpage
{ \Large{Supplementary Materials}}

\section{Interpretation of higher order correlators}

By construction, higher order cumulants (Eqs.~(2)-(3) in the main text) contain information that is not contained in the noise power spectrum (Eq.~(1) in the main text). Multidimensional plots of such correlators potentially contain considerably more detailed description of microscopic physics. Here, we will comment about what can be said about a system with a  Lorentzian shape of the noise power spectrum upon measuring higher order spin correlators at the thermodynamic equilibrium.  

According to results of the main text,  measurements of $C_3$ or $C_4$ can be used to determine nonlinear corrections to the average spin relaxation rate $\gamma$ and the coefficients $D_{ij}$ describing the spin-spin correlator in the nonlinear Bloch-Langevin equation. 
First, consider the nonlinear corrections to the average spin relaxation rate: 
\be
\frac{d \la {\bf S} \ra  }{dt} = {\bf B}\times  \la {\bf S} \ra + \left( \gamma_0+\gamma_2  \la {\bf S} \ra  ^2  \right)  \la {\bf S} \ra .
\label{av1}
\ee
The coefficient $\gamma_2$ can be nonzero due to a spin relaxation channel that involves many-body spin interactions, such as exchange coupling during collisions of atoms or conduction electrons. Such interactions are generally important in hot atomic vapors but they are usually negligible in semiconductor electrons. In  semiconductors with electrons developing a Fermi surface, 
spin relaxation is dominated by spin-orbit coupling effects combined with elastic scatterings on magnetic and non-magnetic impurities. Such spin relaxation mechanisms do not lead to nonlinear corrections to the relaxation rate (\ref{av1}). This is  confirmed by our calculations for magentic disorder in the main text.  Hence, a nonlinear correction $\gamma_2$  indicates, for conduction electrons, the presence of an exotic channel of spin relaxation, e.g., enabled by Coulomb scatterings at the background of strong spin-orbit coupling.
 
Here we note that in order to determine the average spin relaxation rate, formally, one does not need an SNS setup. Standard pump-probe techniques can be used to study relaxation of the spin excitation, as far as only average characteristics are considered. 
However, our results in the main text show that the noise term coefficients $D_1$ and $D_2$, defined in Eq.~(9) of the main text, have independent meaning and also describe nonlinear effects. 

The worked out example in the main text shows that $D_1$ can be nonzero even when $\gamma_2=0$. This illustrates the fact that the information content of the higher order spin correlators does not reduce to merely nonlinear corrections to the average response characteristics. In the case of our model of conduction electrons, nonlinear effects described by $D_1$ can  be traced to the Pauli principle that affects dynamics of fluctuations of even noninteracting electrons. This is in contrast to average characteristics that are not influenced by the Pauli principle. Hence, the nonzero value of the higher order correlators of conduction electrons in the model of scatterings on magnetic impurities that was worked out in the main text is a quantum effect that reflects fermionic nature of conduction electrons. 

Let us now discuss the meaning of the coefficient $D_2$ in Eq.~(9) of the main text. Our worked out example of scatterings on magnetic impurities corresponds to $D_2=0$. Also, the term in the Hamiltonian in the path integral that depends on $D_2$ is invariant under the symmetry described by Eq.~(12) of the main text, similarly to the term describing spin precession in the magnetic field. This suggests that nozero values of $D_2$ correspond to 
the presence of noisy magnetic fields that act simultaneously on all electrons in the mesoscopic region. Such fields do not create dissipative transitions but randomize precession of a spin fluctuation.

Indeed, consider dynamics of a spin fluctuation ${\bf S}$ in a noisy magnetic field:
\be
\dot{\bf S} = {\bf b}(t) \times {\bf S}, \quad \la b_{i}(t) b_{j} (t') = D_2 \delta (t-t').
\label{ds1}
\ee
Let $\delta S_i =  \varepsilon_{ijk} S_k \int_{0}^{\delta t} b_j(t) \, dt $ be the change of the spin density during a small interval $\delta t$ such that $\la \delta  S_i \ra \ll S_i$, where $\varepsilon_{ijk}$ is the Levi-Cevita symbol.  Using the definition of the correlator of the  noisy field, we find
\be
\la \delta S_i \delta S_j \ra =  D_2 \delta t \left(S_iS_j - {\bf S}^2 \delta_{ij} \right),
\label{ds2}
\ee
which coincides with the definition of the nonlinear correction described by the coefficient $D_2$ in Eq.~(9) of the main text.
Thus, observation of nonzero $D_2$ values would indicate that the considered electron system interacts with an external source of a noisy magnetic field, e.g., possibly the hyperfine field of nuclear spins.  

Finally, higher order correlators depend also on the value of the parameter $b_0$ that describes the quartic correction to the free energy function in Eq.~(6) of the main text. This parameter contains information about nontrivial static correlations in the thermodynamic equilibrium.  For example, its relative role can become large near a thermodynamic phase transition to a ferromagnetic state, e.g., in a magnetic semiconductor.

\section{Details of calculations of higher order spin cumulants}
In this supplemental material, we provide calculations and analytical expressions for the 3rd and 4th order spin cumulants.
Corresponding to the general form of the Hamiltonian given in Eq.~(14)  in the main text,  the fourth order cumulant can be calculated using the perturbative technique developed in Ref.~\cite{li2013higher}. For this purpose, we first need to obtain the second order correlators. To the leading order, they are determined  by the quadratic parts of the action:  
\be
 &&{\cal R}_2 =\int dt {\cal H}_2  \\
&&=  \int dt \Big( i\bchi \cdot \bS + i \omega_L \bchi\cdot (\bS \times \hat{y})  + i \gamma_0 \bchi \cdot \bS - D_0 \bchi^2\Big), \nn
\ee
where we assume that the external magnetic field is directed along $\hat{y}$ and $|\bB| = \omega_L$. 
After making a Fourier transformation of $\bchi(t)$ and $\bS(t)$, we can rewrite ${\cal R}_2$  in the following form:
\be
{\cal R}_2 = \sum_{\omega>0} {\bf A}^{\dg}(\omega) \hat{M} {\bf A}(\omega), 
\ee
where ${ A}^{\dg}_j(\omega) = A_j (-\omega)$ with 
\be
{\bf A} (\omega) \equiv (\chi_x(\omega), S_x(\omega), \chi_z(\omega), S_z(\omega))^T,
\ee
and 
\be
\hat{M}=\left(
          \begin{array}{cccc}
            -D_0 & -\omega+i\gamma_0 & 0 & i \omega_L \\
            \omega+i\gamma_0 & 0 & -i \omega_L & 0 \\
            0 & -i \omega_L & -D_0 & -\omega+i\gamma_0 \\
            i\omega_L & 0 & \omega+i\gamma_0 & 0 \\
          \end{array}
        \right).
        \label{eq:M}  \nn \\
\ee
Then the second order correlators, to the leading order,  can  be calculated as
\be
G_{ij}(\omega) &=& \la A_i(\omega) A_j (-\omega) \ra_0 \nn \\
&=& \int {\cal D}\bchi \int{\cal D} \bS ~ A_i(\omega) A_j (-\omega)  e^{{\cal R}_2},
\ee
where $\la... \ra_0 $ means that the averaging is taken over quadratic part of the action. Performing the Gaussian integrations, we find  
\begin{widetext}
\be
&& \hat{G} (\omega) = -\hat{M}^{-1}\nn \\
&&~~~= \left(
          \begin{array}{cccc}
            0 & - \dfrac{\strut  \omega+i\gamma_0 }{\strut B_2(\omega) } & 0 & -\dfrac{\strut i \omega_L}{\strut B_2(\omega)} \\
            \dfrac{\strut \omega - i\gamma_0 }{\strut B_2(-\omega)} & \dfrac{\strut 2D_0 (\gamma_0^2+\omega_L^2+\omega^2) } {\strut B_1(\omega)}  & \dfrac{\strut i \omega_L}{\strut B_2(-\omega)} & \dfrac{\strut 4i D_0 \omega_L\omega}{\strut B_1(\omega) } \\
            0 & \dfrac{\strut i \omega_L }{\strut B_2(\omega)} & 0 & - \dfrac{\strut \omega+i\gamma_0}{\strut B_2(\omega)} \\
          - \dfrac{\strut  i \omega_L } {\strut B_2(-\omega)} & - \dfrac{\strut 4i D_0 \omega_L \omega} {\strut B_1(-\omega) } & \dfrac{\strut \omega- i\gamma_0 } {\strut  B_2(-\omega)} & \dfrac{\strut 2D_0 (\gamma_0^2+\omega_L^2+\omega^2)} {\strut B_1(\omega)} \\
          \end{array}
        \right),  \nn 
\ee
\end{widetext}
with $$ B_1(\omega)=\gamma_0^4+(\omega^2-\omega_L^2)^2+2\gamma_0^2( \omega_L ^2+\omega^2),$$ and  $$B_2(\omega)=\omega^2+2i\gamma_0 \omega-\omega_L^2-\gamma_0^2. $$
In the limit of a large magnetic field, i.e. $\omega_L \gg \gamma_0$,  expressions, for $\omega>0$, simplify:
\be
\hat{G}(\omega) =  \left(
          \begin{array}{cccc}
            0 & -S_2 & 0 & -i S_2 \\
            S_2^* & S_1 &i S_2^*& iS_1 \\
            0 &iS_2 & 0 & -S_2 \\
          -i S_2^* & - i S_1& S_2^* & S_1\\
          \end{array}
        \right),  \nn 
\ee
with $$S_1= \frac{D_0}{(\omega- \omega_L)^2+\gamma_0}, \quad S_2 = \frac{1}{2} \frac{1}{\omega- \omega_L + \gamma_0 }. $$

For the calculation of fourth order cumulants, we need to keep the quartic terms, ${\cal R}_4 = \int dt {\cal H}_4$,  that are  fourth order in terms of $\bchi$ and $\bS$ and treat it as a small perturbation. Then the fourth order cumulants can be calculated as 
\be
C_4(\omega_1, \omega_2) =  \int {\cal D}\bchi \int{\cal D}\bS ~ e^{{\cal R}_2} |S_z(\omega_1)|^2 |S_z(\omega_2)|^2   {\cal R}_4 .  \nn\\
\label{eq:c4cal}
\ee

Since the exponent in (\ref{eq:c4cal}) is quadratic in variables, one can perform the integration  using the Wick's rule by summing over all possible combinations of products of second-order correlators. It turns out that all relevant Feynman diagrams have the same topology of an intersection of four propagators.  
Take, as an example, the term proportional to $D_1$ in Eq.(15) from the main text. It contains $4$ terms that contribute to nonzero 4th order cumulants: $D_1 \bS^2 \bchi^2 = D_1 (S_z^2 \chi_z^2 +S_z^2\chi_x^2 + S_x^2 \chi_z^2 +S_x^2\chi_x^2)$ ($y$-components are excluded because they decouple from the measured variables).  For the second term $D_1 S_z^2 \chi_x^2$, the Wick's rule gives rise to a contribution to the 4th order cumulant:
\begin{widetext}
\be
&&4G_{44}(\omega_1) G_{44}(\omega_2) \Big( G_{14}(\omega_1) G_{14}(\omega_2) + G_{14}(\omega_1)G_{41}(\omega_2) + G_{41}(\omega_1) G_{14}(\omega_2) + G_{41}(\omega_1)G_{41}(\omega_2) \Big) \nn\\
&& + 4G_{41}(\omega_1)G_{14}(\omega_1) G_{44}(\omega_2)G_{44}(\omega_2) + 4G_{41}(\omega_2)G_{14}(\omega_2) G_{44}(\omega_1)G_{44}(\omega_1). \nn
\ee
\end{widetext}

  The full calculation is straightforward.  Each term in the action of the path integral leads to a separate distinct contribut. Hence, the fourth order cumulant is the sum of the following terms: 
  \be
  C_4= C_4^{(1)} +C_4^{(2)}+ C_4^{(3)} + C_4^{(4)} + C_4^{(5)},  
  \ee
  corresponding to, respectively, the terms with coefficients $D'_1 \equiv D_1-D_2$, $D_2$, $D_3$, $D_4$ and $\gamma_2$ in Hamiltonian (Eq.(14)) in the main text. 
  We provide the analytical expressions for two limiting cases. One is with large external magnetic field, i.e. $\omega_L \gg \gamma_0$. 
\be
&&C_4^{(1)} = D'_1 \frac{2D_0^2 (8\gamma_0^2 + (\Omega_1- \Omega_2)^2)}{(\Omega_1+\gamma_0^2)^2 ( \Omega_2^2 + \gamma_0^2)^2} , 
\ee
\be
&&C_4^{(2)} = D_2 \frac{4D_0^2 (4\gamma_0^2 + (\Omega_1 + \Omega_2)^2)}{(\Omega_1+\gamma_0^2)^2 ( \Omega_2^2 + \gamma_0^2)^2} , 
\ee
\be
&&C_4^{(3)} = D_3 \frac{4D_0 \gamma_0 (2\gamma_0^2 + \Omega_1^2 + \Omega_2^2)}{(\Omega_1+\gamma_0^2)^2 ( \Omega_2^2 + \gamma_0^2)^2} , 
\ee
\be
&&C_4^{(4)} = D_4 \frac{4}{(\Omega_1+\gamma_0^2) ( \Omega_2^2 + \gamma_0^2)} ,
 \ee
\be
&&C_4^{(5)} = - \gamma_2 \frac{32 D_0^3 \gamma_0 }{(\Omega_1+\gamma_0^2)^2 ( \Omega_2^2 + \gamma_0^2)^2}  .
\ee
where we defined $\Omega_{1, 2} = \omega_{1,2} - \omega_L$. 

The other case is without external magnetic field, i.e. $\omega_L=0$:
\be
&&C_4^{(1)}   = D'_{1} \frac{ 16 D_0^2 (6\gamma_0^2 + \omega_1^2 + \omega_2^2)}{(\omega_1+\gamma_0^2)^2 ( \omega_2^2 + \gamma_0^2)^2},
  \ee
  \be
&&C_4^{(2)}   = D_{2} \frac{ 16 D_0^2 (6\gamma_0^2 + \omega_1^2 + \omega_2^2)}{(\omega_1+\gamma_0^2)^2 ( \omega_2^2 + \gamma_0^2)^2},
  \ee
\be
&&C_4^{(3)}   = D_{3} \frac{ 24 D_0 \gamma_0 ( 2\gamma_0^2 + \omega_1^2 + \omega_2^2)}{(\omega_1+\gamma_0^2)^2 ( \omega_2^2 + \gamma_0^2)^2} ,
 \ee
 
\be
&&C_4^{(4)} = D_4 \frac{24}{(\omega_1+\gamma_0^2) ( \omega_2^2 + \gamma_0^2)}  ,
\ee

\be
 &&C_4^{(5)} = -\gamma_2 \frac{192 D_0^3 \gamma_0 }{(\omega_1+\gamma_0^2)^2 ( \omega_2^2 + \gamma_0^2)^2} .
\ee

When there is external pumping, which breaks time-reversal symmetry,  the third order cumulant is nonzero. Its calculation follows a similar procedure as for the 4th order one. It is the sum of the following terms
\be
  C_3= C_3^{(1)} +C_3^{(2)}+ C_3^{(3)} + C_3^{(4)} + C_3^{(5)},  
  \ee
  corresponding to, respectively, the terms with coefficients $D'_1 \equiv D_1-D_2$, $D_2$, $D_3$,  $\gamma_2$ and $k_p$ in the Hamiltonian (Eq.(20)) in the main text. 
 For the limiting case with $\omega_L \gg \gamma_0$: 
\be
 C_3^{(1)} = 2D'_1 \frac{ C_S D_0 \Big( 2\gamma_0^2 + (\Omega_1 +\Omega_2)^2 -i\gamma_0(\Omega_1 + \Omega_2 )\Big)}{ (\Omega_1^2 + \gamma_0^2) (\Omega_2^2 + \gamma_0^2 ) [(\Omega_1 + \Omega_2 )^2 +\gamma_0^2]} , \nn
\ee
\be
C_3^{(2)} = D_2 \frac{C_S  D_0 \Big( 4\gamma_0^2 + \Omega_1^2 +\Omega_2^2 + i\gamma_0(\Omega_1 + \Omega_2 )\Big)}{ (\Omega_1^2 + \gamma_0^2) (\Omega_2^2 + \gamma_0^2 ) [(\Omega_1 + \Omega_2 )^2 +\gamma_0^2]}, \nn
\ee
\be
C_3^{(3)} = i D_3 \frac{ C_S }{ (\Omega_1+ i\gamma_0)(\Omega_2 + i\gamma_0)(\Omega_1+\Omega_2 - i \gamma_0)}, \nn
\ee
\be
&& C_3^{(4)} = -\gamma_2  \frac{12 C_S  D_0^2 \gamma_0 }{ (\Omega_1^2 + \gamma_0^2) (\Omega_2^2 + \gamma_0^2 ) [(\Omega_1 + \Omega_2 )^2 +\gamma_0^2]} , \nn
\ee
\be
C_3^{(5)} = \frac{k_p}{4} \frac{D_0 (3\gamma_0^2 + \Omega_1^2 + \Omega_1\Omega_2 +\Omega_2^2)}{ (\Omega_1^2 + \gamma_0^2) (\Omega_2^2 + \gamma_0^2 ) [(\Omega_1 + \Omega_2 )^2 +\gamma_0^2]}  , \nn
\ee
with $C_S \equiv \bar{S}_z + i \bar{S_x}$.

For the case without external magnetic field, $\omega_L =0 $:
\be
&&C_3^{(1)} = D'_{1} \frac{ 4  \bar{S}_z D _0 (3\gamma_0^2 + \omega_1^2 + \omega_1\omega_2 + \omega_2^2)}{ (\omega_1^2 + \gamma_0^2) (\omega_2^2 + \gamma_0^2 ) [(\omega_1 + \omega_2 )^2 +\gamma_0^2]}  , \nn
\ee

\be
&&C_3^{(2)} = D_{2} \frac{ 4  \bar{S}_z D _0 (3\gamma_0^2 + \omega_1^2 + \omega_1\omega_2 + \omega_2^2)}{ (\omega_1^2 + \gamma_0^2) (\omega_2^2 + \gamma_0^2 ) [(\omega_1 + \omega_2 )^2 +\gamma_0^2]}  , \nn
\ee

\be
&&C_3^{(3)} = i D_3 \frac{ 6 \bar{S}_z }{ (\omega_1+ i\gamma_0)(\omega_2 + i\gamma_0)(\omega_1+\omega_2 - i \gamma_0)}  , \nn 
\ee

\be
&& C_3^{(4)} = -\gamma_2  \frac{ 72  \bar{S}_z  D_0^2 \gamma_0 }{ (\omega_1^2 + \gamma_0^2) (\omega_2^2 + \gamma_0^2 ) [(\omega_1 + \omega_2 )^2 +\gamma_0^2]} ,\nn 
\ee

\be
&&C_3^{5} = k_p \frac{2 D_0 (3\gamma_0^2 + \omega_1^2 + \omega_1\omega_2 +\omega_2^2)}{ (\omega_1^2 + \gamma_0^2) (\omega_2^2 + \gamma_0^2 ) [(\omega_1 + \omega_2 )^2 +\gamma_0^2]} .\nn\\
\ee

\bibliographystyle{apsrev4-1}
%\bibliography{bib_sns}

%%%%%%%%%%%%%%%%%%%%%%%%%%%%%%5%%%%%%%%%%%%%%%%%%%%
%%%%%%%%%%%%%%%%%%%%%%%%%%%%%%%%%%%%%%%%%%%%%%%%%%%%%
\end{document}